\documentclass{ws-ijmpa}

\begin{document}

\markboth{Volker D. Burkert}
{Highlights of recent results with CLAS}

%
\catchline{}{}{}{}{}
%

\title{\bf HIGHLIGHTS OF RECENT RESULTS WITH CLAS\\}

\author{\footnotesize Volker D. Burkert}

\address{Physics Division, Jefferson Lab, 12000 Jefferson Avenue\\
Newport News, VA 23606,
USA}

\maketitle
\centerline{For the CLAS Collaboration}
\vspace{0.3cm}
\begin{abstract}
Recent results on the study of the electromagnetic structure of nucleon resonances, the spin structure of protons and neutrons at small and intermediate photon virtuality, and the search for exotic pentaquark baryons  are presented. 

\end{abstract}

\section{Introduction}	

One of the central questions in hadronic physics is what are the relevant degrees of 
freedom when probing nucleons at different distance scales? The experimental program at 
Jefferson Lab and measurements using the CLAS detector, are aimed at providing answers 
to this fundamental question. In this presentation I will report on recent results obtained 
with CLAS. Longstanding issues that are now 
being addressed experimentally, as well as theoretically, include the study of electromagnetic 
transition form factors, the baryon resonance excitation spectrum, and the spin 
structure of the nucleon in the transition from small to large distances. Another, highly 
relevant subject is the search for pentaquark baryons that are claimed to have been observed 
recently in about a dozen experiments.     

\section{The nucleon spin structure in the transition regime.}
Understanding the spin structure of 
the nucleon in the transition from the single parton regime through multiparton effects 
(higher twists) and the regime of nucleon resonances is an ongoing challenge to the nuclear 
physics community. Measurements of inclusive structure functions and their moments are 
important tools towards this goal. Also, recent theoretical developments show that single 
spin asymmetries for semi-inclusive deep inelastic scattering (SIDIS) can probe orbital 
angular momentum contributions to the nucleon spin. CLAS has a large program to measure the spin 
structure function $g_1(x,Q^2)$ for hydrogen and deuterium for a broad range of energies 
from 1.6 GeV to 5.75 GeV. At the same time, polarized target asymmetries for semi inclusive 
meson production are obtained.
\begin{figure}[t]
\vspace*{6.5cm}
\centering{\includegraphics{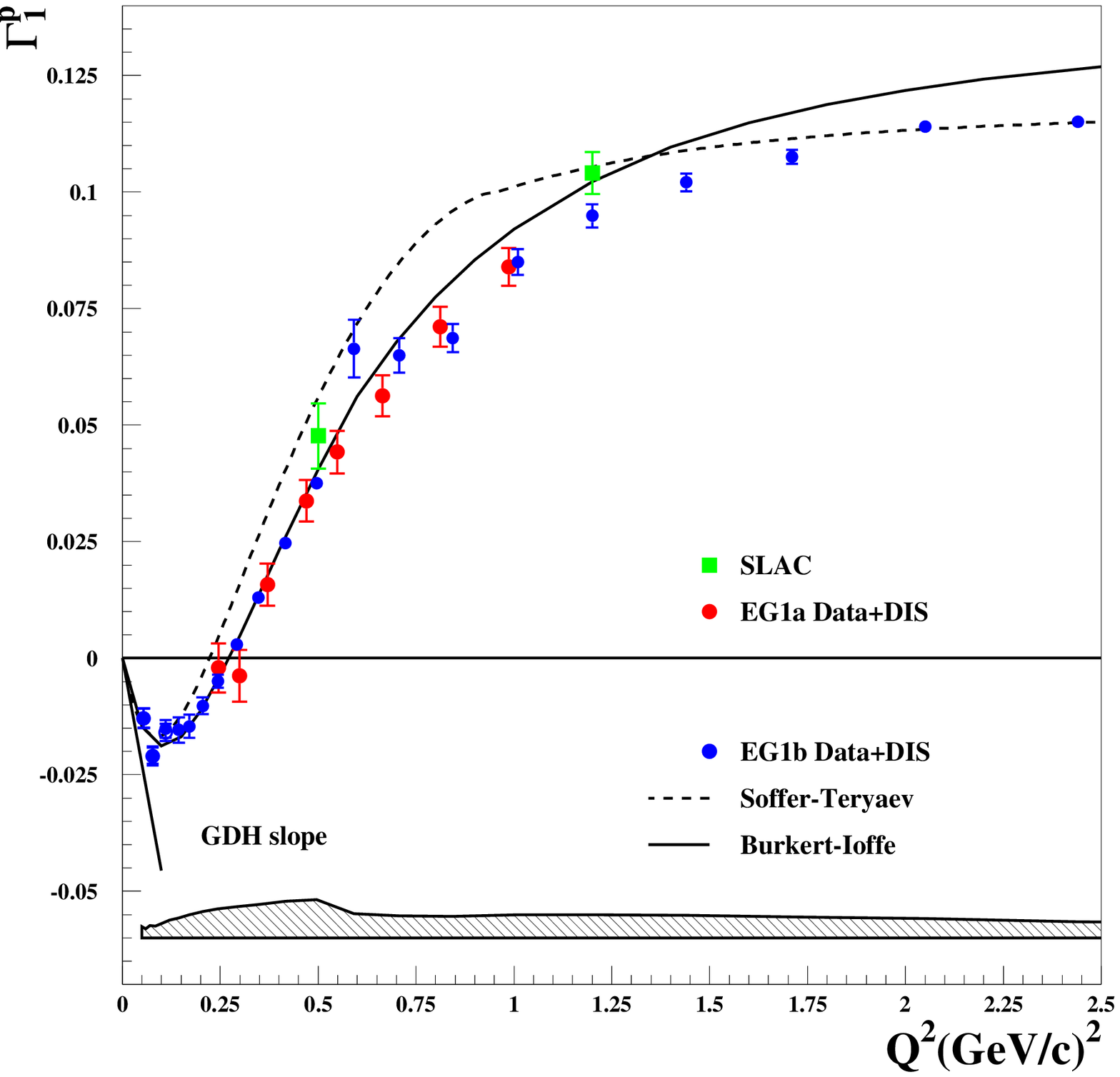}}
\centering{\includegraphics{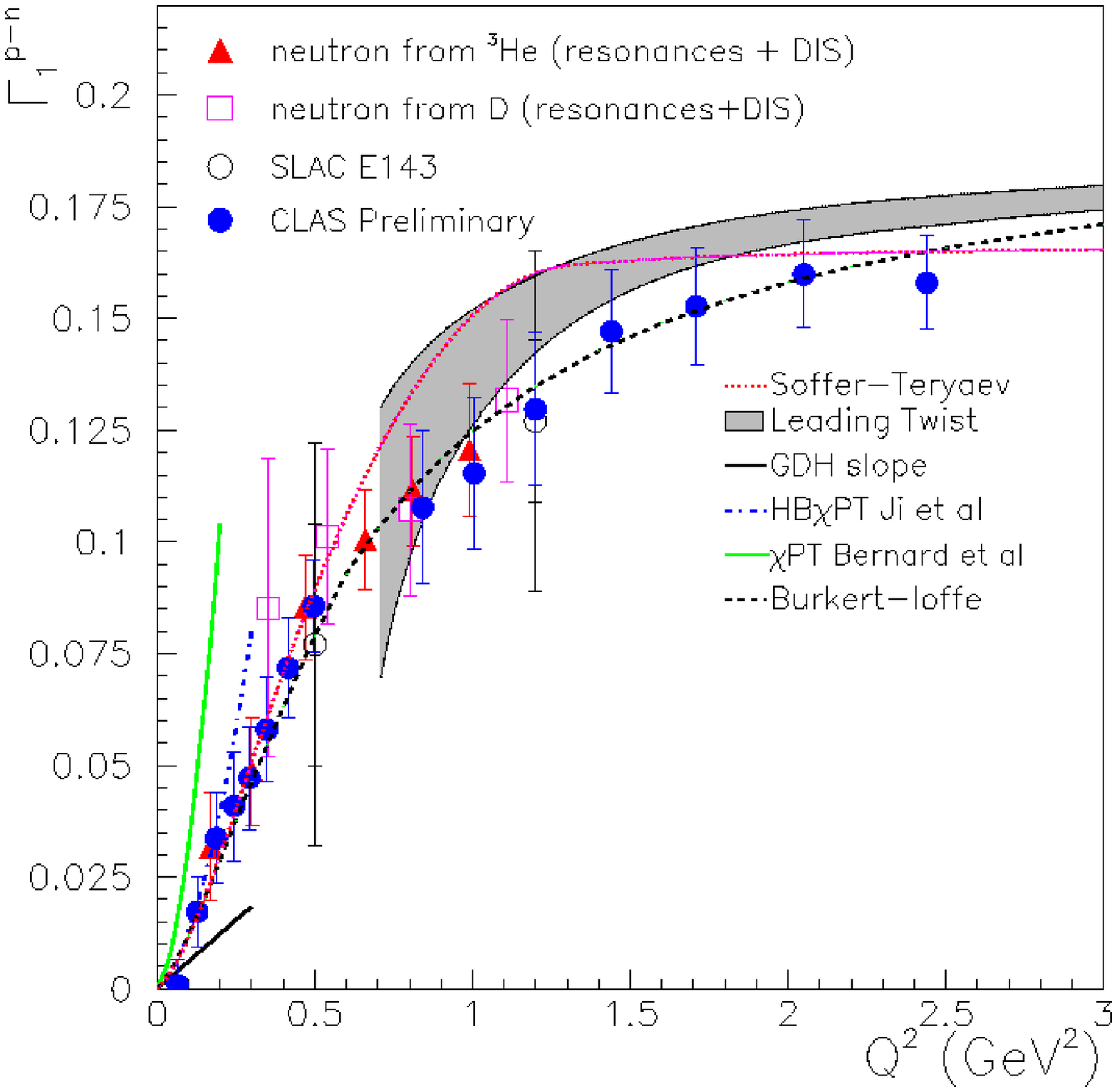}}
\caption[]{\small{Left panel: The first moment $\Gamma_1^p(Q^2)$ for the proton. The full circles are CLAS data. 
 Right panel: The Bjorken integral. The full triangles represent
results extracted from published data from CLAS and Hall A data\cite{deur04}, and the open squares 
are from published CLAS results. The full circles represent preliminary CLAS data.}}
\label{fig:gamma1}
\end{figure} 

\subsection{Inclusive polarization asymmetries and sum rules}
There is significant interest in  the spin structure function $g_1(x,Q^2)$ and the first moment $\Gamma_1=\int_0^{1}g_{1}(x,Q^2)dx$, both for the proton and neutron, as well as for the p-n difference. The latter is directly related to the Bjorken sum rule, $\Gamma_1^p-\Gamma_1^n = \frac{g_A}{6}$, at asymptotic values of $Q^2$. At finite $Q^2$ these moments probe contributions of multiple parton effects (higher twist), resonance excitations, and other coherent effects.
The slopes of the integrals at $Q^2=0$ are directly related to the Gerasimov-Drell-Hearn sum rule (GDH-SR). 
The results of a first determination of $\Gamma_1(Q^2)$ was based on measurements at energies of 
2.5 and 4.2 GeV, for both proton\cite{fatemi03} and deuteron\cite{yun03}. New preliminary results include 
high statistics measurements at 1.6 and 5.7 GeV and extend the kinematics range and 
statistical accuracy considerably. The first moment $\Gamma_1^p(Q^2)$ and the Bjorken integral $\Gamma_1^{p-n}$ are shown in Fig.~\ref{fig:gamma1}. In these results the unmeasured part of the DIS contribution has been 
added using parameterizations of results obtained in high energy experiments. $\Gamma_1^p(Q^2)$ shows a 
strong $Q^2$ dependence below $Q^2=1.5$~GeV$^2$ with a zero crossing near $Q^2 = 0.3$~GeV$^2$. At very small $Q^2$, the data approach the 
slope given by the GDH-SR. Similar data sets have been obtained for deuterium. By combining data from
deuterium and hydrogen, $\Gamma_1^n(Q^2)$ for the neutron and the Bjorken integral have been determined in 
the range $Q^2 = 0.06 - 2.5$~GeV$^2$. In the upper $Q^2$ regime the results are consistent with the 
operator product expansion of pQCD in leading twist, and the low $Q^2$ part is described by Heavy Baryon Chiral Perturbation Theory\cite{xji}.
While the Bjorken integral may be most suitable for a consistent description\cite{burkert01}, there is currrently no 
formulation of the entire $Q^2$ regime based on fundamental theory. 
However, various model descriptions have been attempted\cite{ioffe1,ioffe2,soffer}. 
A good description of the transition region is obtained in a vector meson picture that also includes 
parameterizations of s-channel nucleon resonance excitations\cite{ioffe2}.

\subsection{Single spin asymmetries}
\begin{figure}[t]
\vspace*{4.5cm}
\centering{\includegraphics{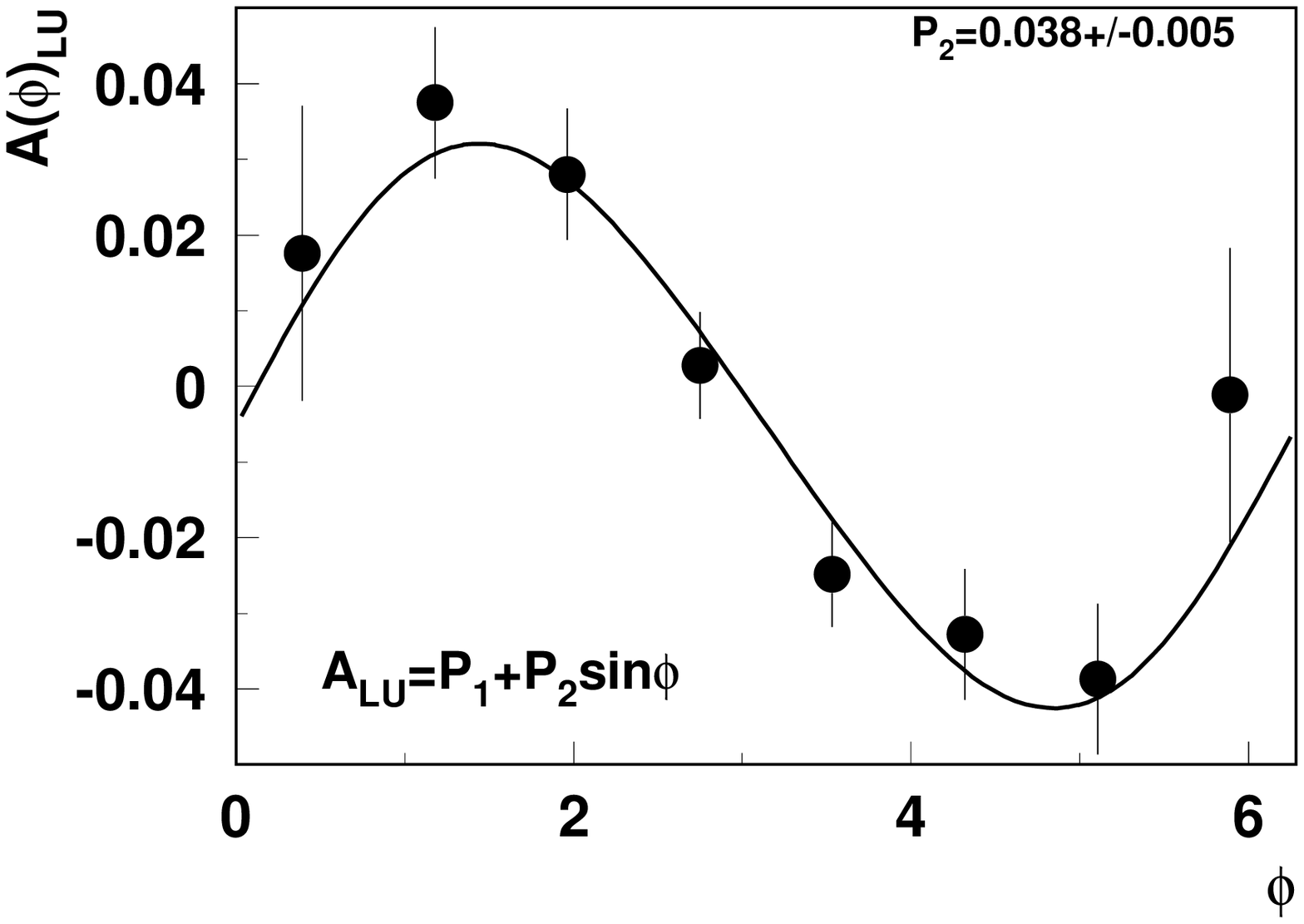}}
\centering{\includegraphics{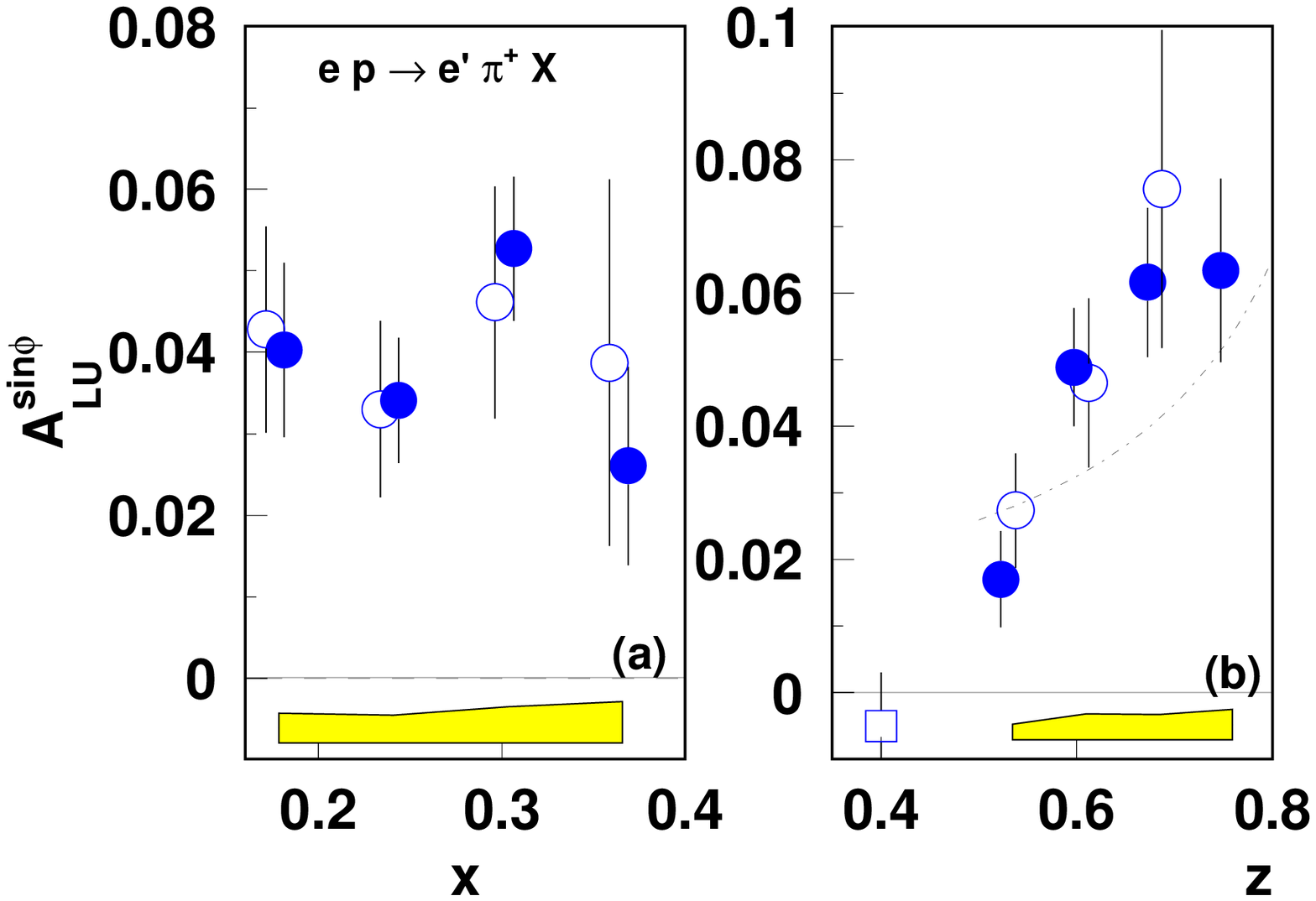}}
\caption[]{\small {Beam-spin azimuthal asymmetry from CLAS as a function of $\phi$ (left panel). The two right panels show the dependence of the $\sin\phi$ moment $A_{LU}^{\sin\phi}$ on $x$ in the range $0.5 < z < 0.8$, and as a function of $z$, in the ranges $0.15 < x < 0.4$ for $M_X > 1.1$GeV (filled circles). The empty circles show the same but for $M_X > 1.4$ GeV. The empty square shows the HERMES result\cite{hermes_ssa}
, which is an average over the range $z$=0.2-0.7 and $x$=0.02-0.4. } }
\label{fig:ssa}
\end{figure}

Measurements of single spin asymmetries have become another focus of hadronic physics during the past few years. A much improved theoretical understanding has been obtained\cite{efremov}, and their relation to the transverse motion of quarks is currently under intense investigation. For the first time a non-zero azimuthal beam spin asymmetry in $ep\rightarrow e^{\prime}\pi^+X$~\cite{avakian04} was measured at CLAS. The data are shown in Fig.~\ref{fig:ssa}. These results also showed that semi-inclusive data taken at 4.2 GeV beam energy in DIS kinematics are consistent with a parton picture interpretation\cite{avakian04}. The $x$ and $z$ dependence of the asymmetry are presented in Fig.~\ref{fig:ssa}. The asymmetry $A_{LU}$ was shown to be directly linked to the twist-3 structure function $e(x)$, which is related to the nucleon $\Sigma_N$ term\cite{efremov}. Much more data have been collected using polarized beams and longitudinally polarized targets at higher beam energies. For example, the target asymmetry $A_{UL}$ has been measured for $\pi^{\pm,0}$ production in a large kinematics range\cite{avakian_gdh04}.  
 
\section{Nucleon resonances.}

\subsection{The $N\Delta$ transition form factors.}
\begin{figure}[t]
\vspace*{6cm}
\centering{\includegraphics{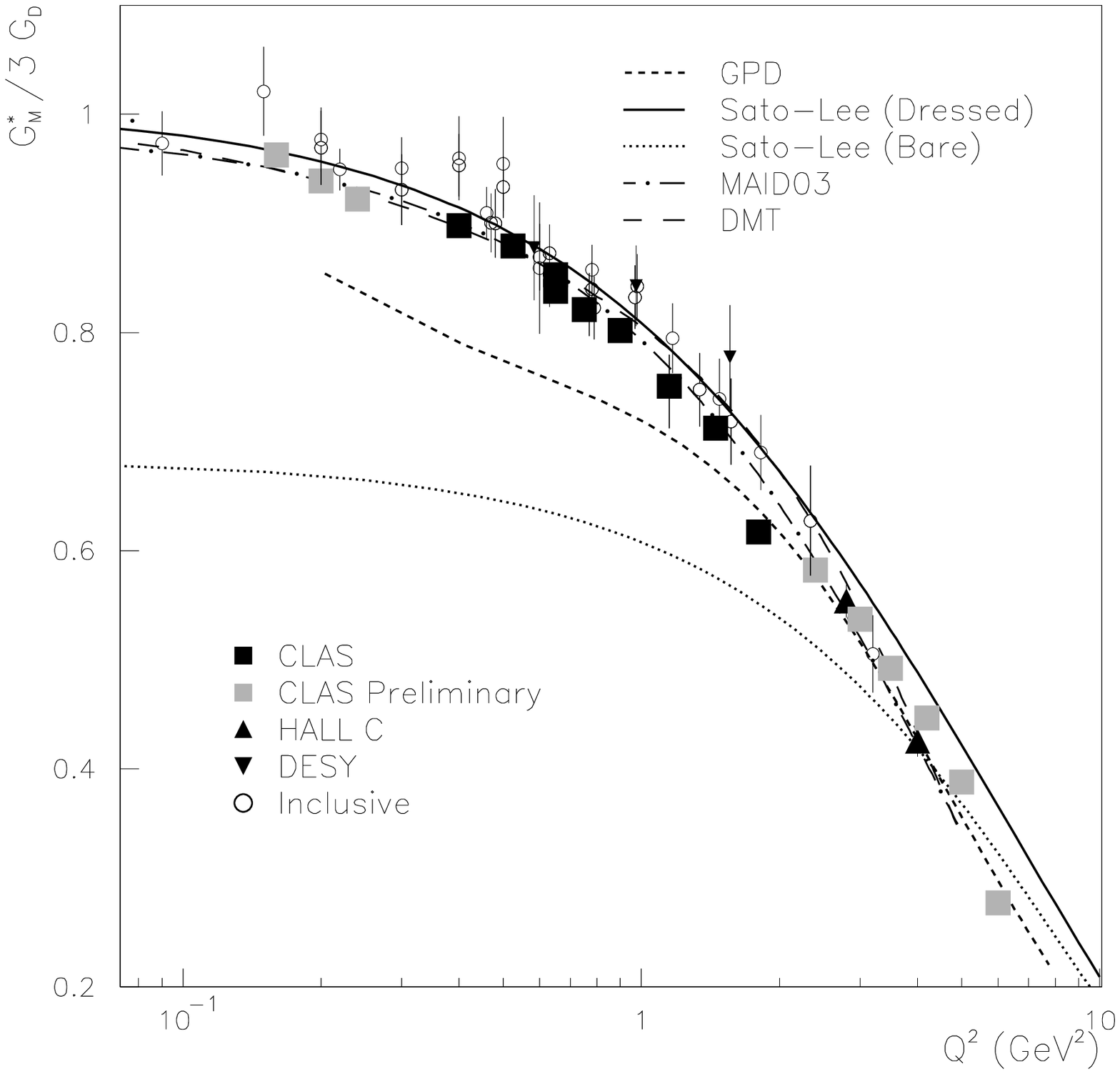}}
\centering{\includegraphics{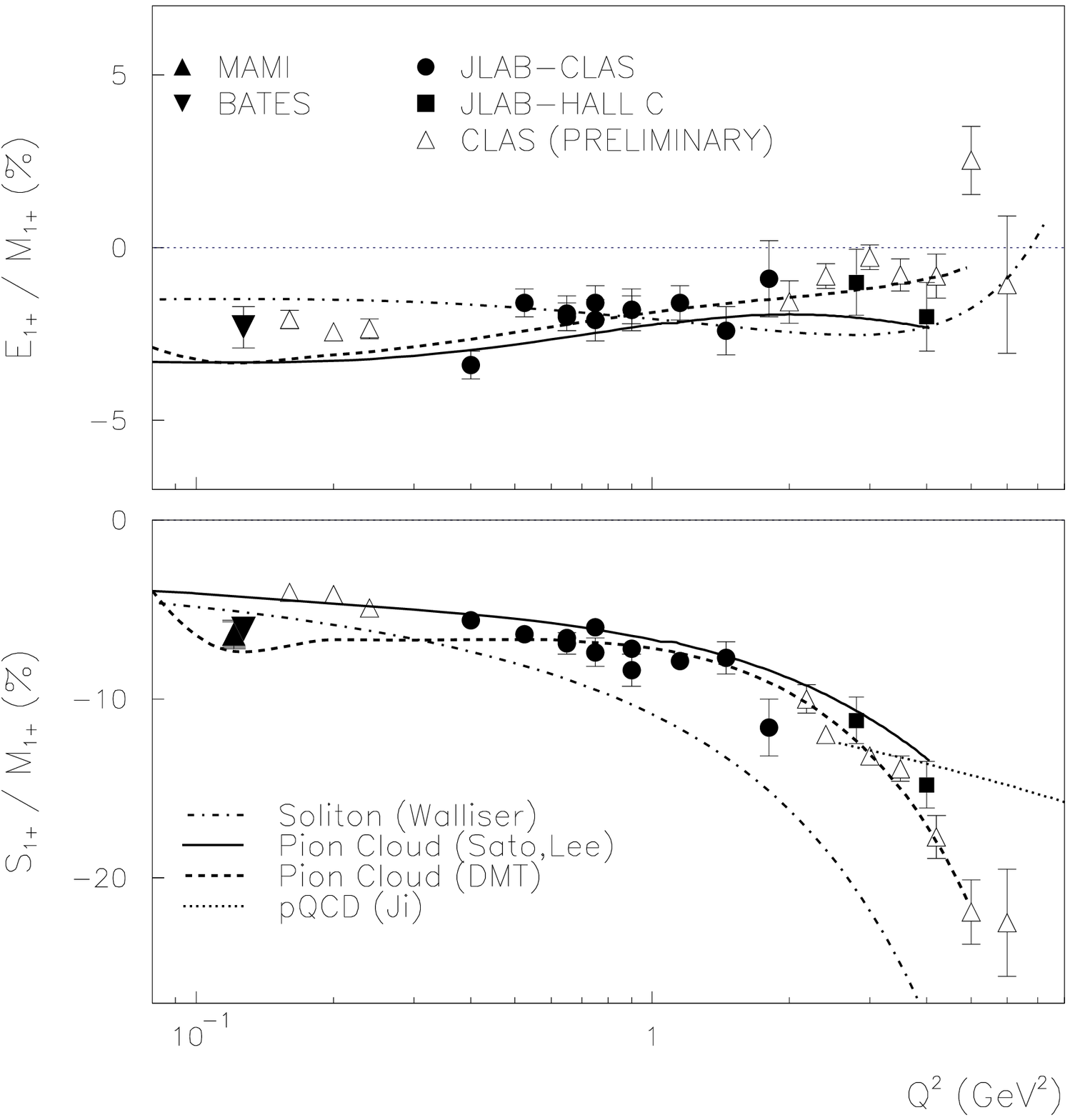}}
\caption{\small Left panel: The magnetic $N\Delta$ transition form factor $G_M^{\Delta}$ determined from the $M_{1+}$ multipole. 
Right panel: Multipole ratios $R_{EM}$ and $R_{SM}$ vs. photon virtuality $Q^2$. The systematic uncertainties for the preliminary $R_{EM}$ 
and $R_{SM}$ data are currently $\pm 2$\% absolute, but should be significantly reduced in the future. }
\label{fig:gm1}
\label{fig:delta_multipoles}
\end{figure}

An interesting aspect of nucleon structure studies at low and intermediate energies is the possible deviation of the $\Delta(1232)$ shape from spherical symmetry. In spherically symmetric $SU(6)$ models, the only contribution is the $M_{1+}$ magnetic dipole transition.  Dynamically, deviations from spherical symmetry may result from interactions of the virtual photon with the nucleon pion cloud, or from one-gluon exchange contributions at small distances, and from genuine D-state contributions in the wave function. In model interpretations\cite{buchmann01,buchmann02} non-spherical components to the wave function give rise to non-zero values of the quadrupole transition multipole $E_{1+}$ for the $N\Delta$ transition. Recent calculations in quenched and full lattice QCD\cite{alexandrou03,alexandrou04} are in support of these model estimates. At asymptotically short distances, i.e. $Q^2 \to \infty$, a model-independent prediction requires $R_{EM}=E_{1+}/M_{1+} \to +1$ and $R_{SM} = S_{1+}/M_{1+} \to constant$. The interpretation of these ratios in terms of quadrupole deformation can therefore only be valid at relatively low momentum transfer. Accurate measurements of $R_{EM}$ and $R_{SM}$ at low and intermediate $Q^2$ are crucial in these studies. The CLAS collaboration has already published cross sections and complete angular distributions of $\pi^0$ production from protons\cite{kjoo02}, and extracted the multipole ratios $R_{EM}$ and $R_{SM}$ in the $Q^2$ range of $0.4 - 2$~GeV$^2$. These measurements have now been extended to lower as well as to higher $Q^2$ values. Preliminary results for the magnetic transition form factor $G_M^*$, and for the multipole ratios are shown in Fig.~\ref{fig:delta_multipoles}. $R_{EM}$ remains negative and small throughout the $Q^2$ range covered.  The multipole ratios agree in sign and magnitude with the aforementioned LQCD calculations. Interpretations within LQCD and dynamical models consistently give a small oblate deformation of the $\Delta(1232)$ from spherical symmetry.

\subsection{The Roper $P_{11}(1440)$, $S_{11}(1535)$. }
\begin{figure}[bt]
\vspace*{8.5cm}
\centering{\includegraphics{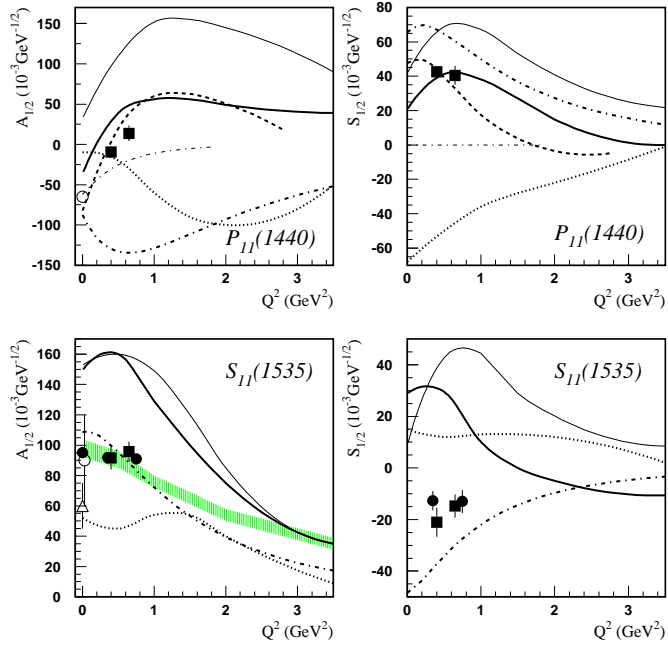}}
\caption[]{\small {Transverse (left) and longitudinal (right) helicity amplitudes for the 
$\gamma pP_{11}(1440)$ transition (upper panels), and 
the  $\gamma pS_{11}(1535)$ transition (lower panels). Bold and thin solid lines correspond to 
relativistic and non-relativistic quark model calculations\cite{capstick95}. 
Dashed lines correspond to light-front calculations\cite{pace99}. 
Dotted, dashed-dotted and thin dashed lines correspond to quark models \protect\cite{warns90,aiello98,cano98}.}}
\label{fig:p11_s11}
\end{figure}
The second nucleon resonance region contains 3 isospin $\frac{1}{2}$ states, the Roper $P_{11}(1440)$, the $S_{11}(1535)$, and the $D_{13}(1520)$. The increased complexity of the partial waves contributing to single pion production requires measurement of other isospin channels, and inclusion of polarization observables in the analysis. Unitary models have been used in the extraction of the resonance contributions\cite{burkert_lee_04}. Recent CLAS data on $p\pi^0$ and $n\pi^+$ cross section and beam symmetry data have been analyzed to extract resonance contributions to these nucleon states using a unitary isobar model as well as a dispersion relation approach\cite{aznauryan04}. The $S_{11}(1535)$ has also been analyzed using the $p \eta$ differential cross section data. The results for two $Q^2$ points are shown in Fig.~\ref{fig:p11_s11}. 

The most surprising result is seen for the $P_{11}(1440)$. The transverse photocoupling amplitude $A_{1/2}(Q^2)$ drops rapidly in magnitude with $Q^2$, and appears to change sign near $Q^2 = 0.5$~GeV$^2$. The longitudinal amplitude, which was long considered consistent with zero, is large and positive. Simple non-relativistic quark models are unable to describe the data. The best agreement is obtained by a model that constructs the Roper as a state with a small quark core and a large meson cloud\cite{cano98}. It correctly predicts the sign change as well as sign and magnitude of the longitudinal amplitudes $S_{1/2}$. Models that include relativity provide a better description than non-relativistic approaches.             

The photocoupling amplitudes for the $S_{11}(1535)$ are also well determined for these low $Q^2$ points. The unitary isobar model and the dispersion relation approach give consistent results. Moreover, the analysis of the $N\pi$ and the $p\eta$ channels are in good agreement. These new results do not support earlier analyses of photoproduction data that showed significantly different results for the $p\eta$ and the $N\pi$ channels. 

The $D_{13}(1520)$ amplitudes are also well determined. At the real photon point, the $A_{3/2}$ amplitude is dominant. At the $Q^2=0.5-0.6$~GeV$^2$, the two amplitudes are about equal in magnitude, and opposite in sign. This confirms the earlier observation of a rapid change in the helicity structure of this state, from helicity $\frac{3}{2}$ dominance at the photon point to helicity $\frac{1}{2}$ dominance at high $Q^2$. 

\subsection{Search for ``missing'' baryon states}

It is well known that the symmetric quark model predicts many states with masses below 2 GeV that have not been found experimentally. This has been called the problem of the "missing resonances". The absence of states that are predicted in the symmetric model may have a fundamental origin, namely that the basic symmetry group underlying the baryon spectrum is not $SU(6)\otimes O(3)$. This could have important implications for the internal baryon structure, and other symmetry groups have recently been studied\cite{kirchbach}. A much simpler explanation could be, that the missing states just haven't been found yet. Indeed most of the known states have been observed in $\pi N$ scattering. If a state couples weakly to $\pi N$, it is not surprising that it hasn't been seen. This conjecture is supported by calculations of decay properties of baryon resonances\cite{koniuk,caprob}, and many of the unobserved states are indeed predicted to have small $\pi N$ widths. How can we search for these states? Many of the ``missing'' states are predicted to have strong $N\pi\pi$ couplings, and some with significant $p\omega$ or $K\Lambda$ widths. A number of states are also predicted with significant $\gamma N$ coupling. This has led to systematic searches for "missing" states in the channels $\gamma p \rightarrow p\pi^+\pi^-$, $\gamma p \rightarrow p\pi^0\pi^0$, $\gamma p \rightarrow \rightarrow K^+\Lambda$, $\gamma p \rightarrow p\omega$, and others. While many hints of new resonant states are clearly present in these data sets, no new state has yet been convincingly identified with its partial wave content. Development of more sophisticated analysis procedures to deal with complex background amplitudes are needed to make further progress. In the following I will discuss the $p\pi^+\pi^-$ channel where new data have been analyzed\cite{ripani03,bellis04}.     

\begin{figure}[bt]
\vspace*{5.5cm}
\centering{\includegraphics{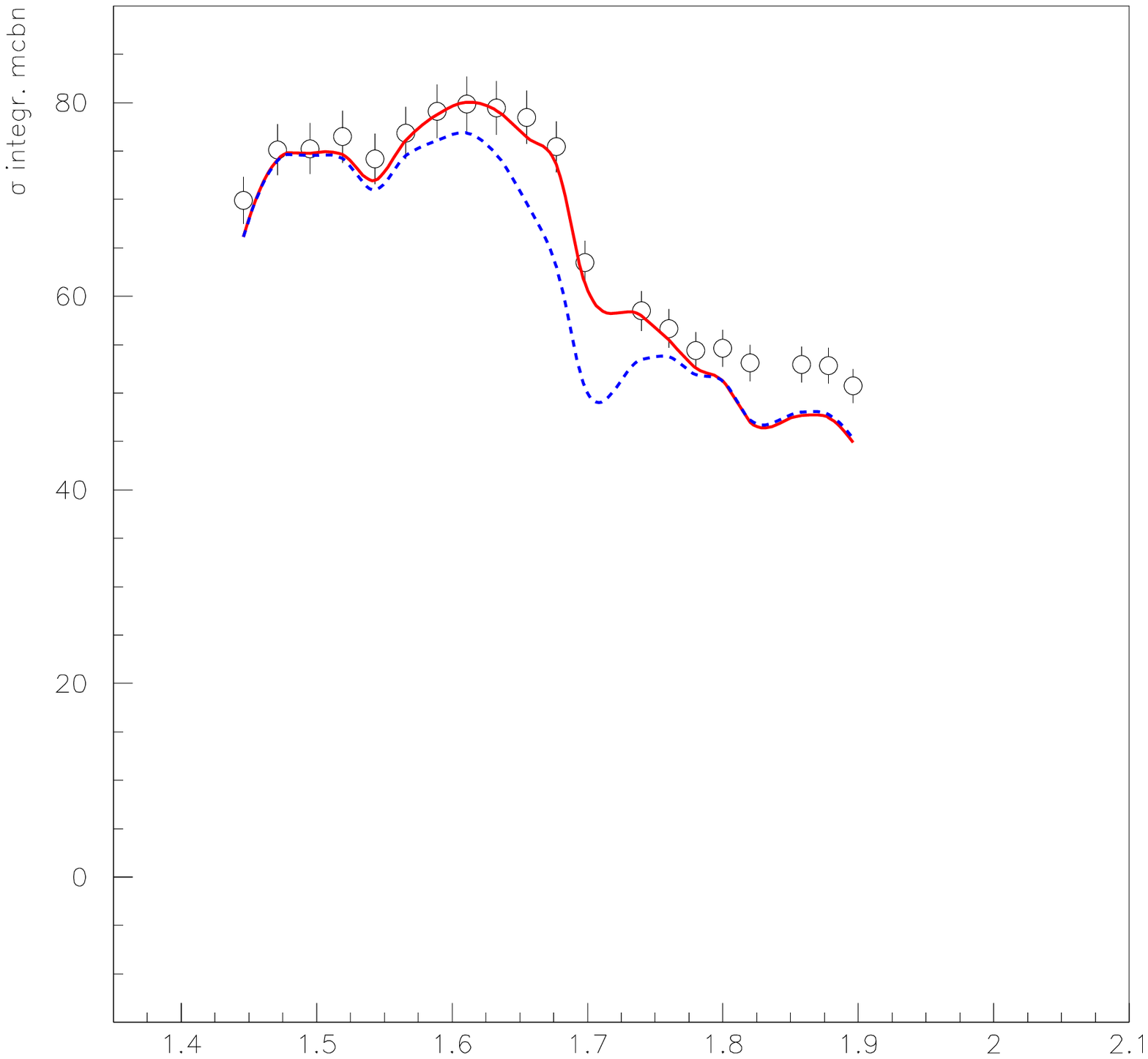}}
\centering{\includegraphics{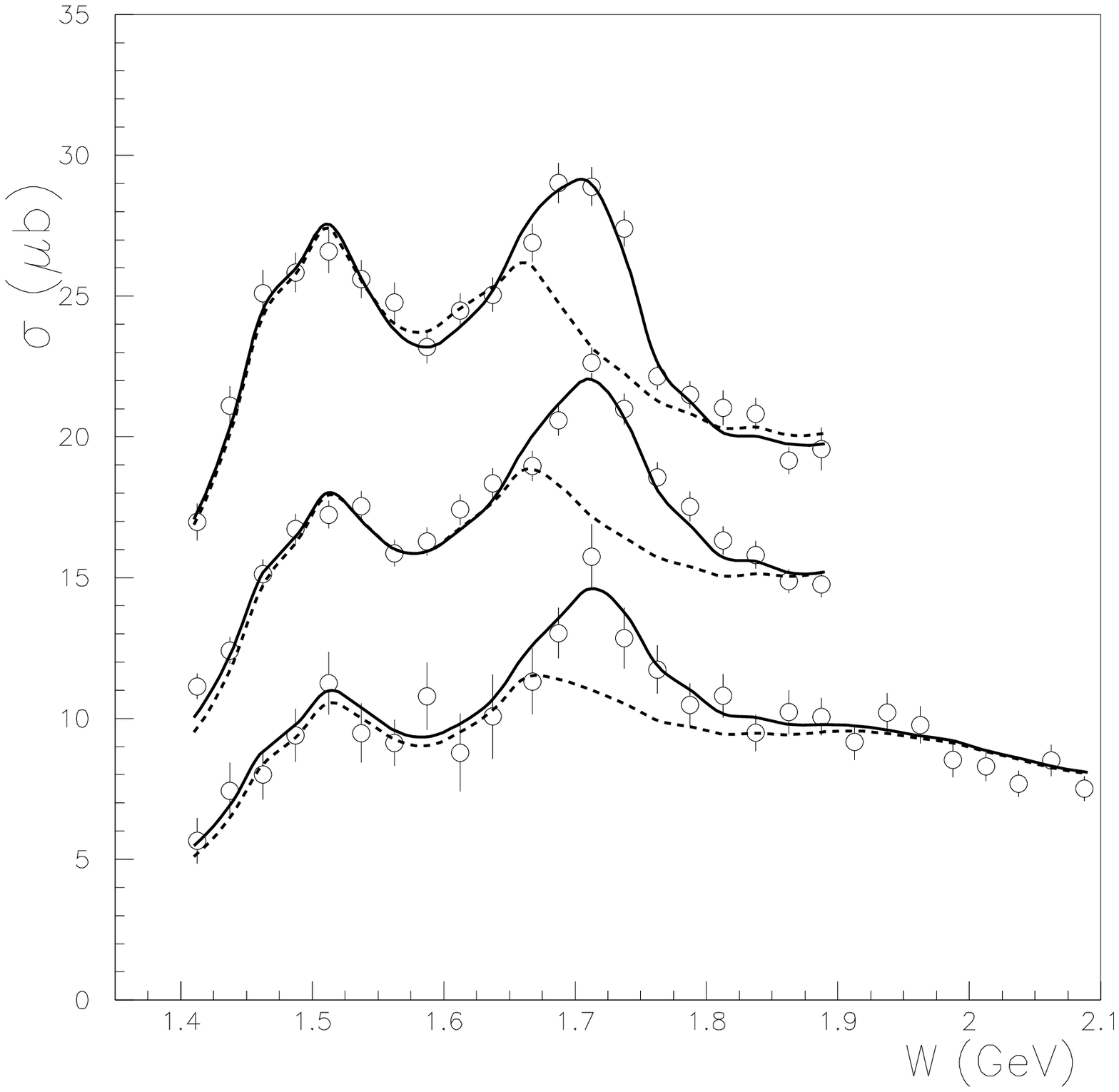}}
\caption{\small {Total cross section for $p\pi^+\pi^-$ in photoproduction (left panel), and electroproduction (right panel) at three $Q^2$ values, 0.65, 0.95, 1.3 GeV$^2$. 
The curves are described in the text.} }
\label{fig:cs_ppi+pi-}
\end{figure}

Analysis of photoproduction of charged two-pion production is complicated by large background contributions besides the resonant amplitudes. For example, the channel $\gamma p \rightarrow p \rho$ has large background contributions due to diffractive scattering, especially if the initial photon is "real". This contribution will dominate the forward angle region. With increasing photon virtuality, $Q^2$, this background component should decrease, which makes electroproduction possibly a more sensitive probe for resonances in that channel. Analysis procedures can be divided into two distinctly different approaches. A frequently used approach is based on an energy-dependent description of the reaction within an isobar model, which is used in this analysis. The energy-dependence of the amplitudes is parametrized as a sum of resonant and non-resonant amplitudes. Non-resonant amplitudes are described by the Born terms for the isobar sub-channels and other contributions. Resonances are parameterized as relativistic Breit-Wigner forms. Fits are done to one-dimensional projections of the cross section on invariant mass distributions and angular distributions. To search for new resonances, s-channel Breit-Wigner forms are introduced in some partial waves and included in the fit. This approach makes maximum use of accumulated knowledge from hadronic processes, and can provide a good description of the projected data. The method has been used in the analysis of CLAS electroproduction data\cite{ripani03}. In this analysis a significant discrepancy between the data and the resonance parameterizations implemented in the fit model was found near a mass of 1.7 GeV for the $p\pi^+\pi^-$ system. This was attributed to either inaccurate hadronic couplings for the well known $P_{13}(1720)$ determined in previous analysis of hadronic experiments, or to an additional resonance with $J^P = \frac {3}{2}^+$ with either $I=\frac{1}{2}$ or $I=\frac{3}{2}$.  The discrepancy is more easily seen in the total cross section for electroproduction, shown in Fig.~\ref{fig:cs_ppi+pi-}.  The dotted lines show the model predictions using resonance parameters from single pion electroproduction and from the analysis of $\pi N \rightarrow N\pi\pi$. The solid line represents the fit when the hadronic couplings of the $P_{13}(1720)$ to $\Delta\pi$ and $N\rho$ are allowed to vary significantly beyond the ranges established in the analysis of hadronic data. Alternatively, a new state was introduced with hadronic couplings extracted from the fit while keeping the parameters of the known $P_{13}(1720)$ at the previously established values. In either case, the fit requires a resonance with hadronic couplings that are significantly different from the ones of  $P_{13}(1720)$ listed by PDG. 

Another surprising aspect is that the total photoproduction cross section in the left panel of Fig.~\ref{fig:cs_ppi+pi-} shows a $W$ dependence that is very different from the electroproduction data in the right panel. The peak at 1.72 GeV is completely invisible in the left panel. This is largely due to a much higher background contribution from diffractive $p\rho^0$ production at the real photon point and destructive $N^*$-background interferences. The photoproduction data require a strong resonance near $W = 1.72$ GeV in the $P_{13}$ partial wave which can be identified with the $P_{13}(1720)$~\cite{mokeev04}. In addition, another resonance near 1.72 GeV in the $\frac{3}{2}^+$ partial wave is required with hadronic couplings as extracted from the additional state needed in the 
electroproduction data. It appears that the excitation strength of the PDG state drops faster with $Q^2$ than the strength of the additional $\frac{3}{2}^+$ state, a signature that could be used to discriminate between the two contributions.

\begin{figure}[b]
\vspace{4.5cm}
\centering{\includegraphics{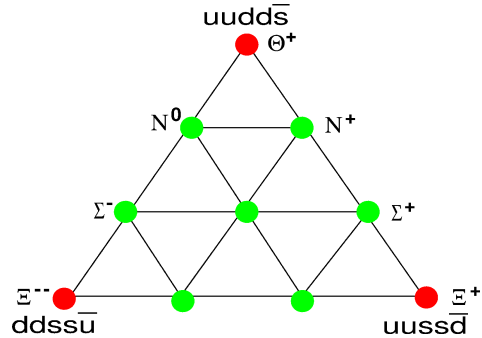}}
\caption{\small{The  ${\bar{10}}$ as predicted in the $\chi SM$}}
\label{fig:antidecuplet}
\end{figure}

\section{Flavor exotic baryons - Pentaquarks } 

If we scatter $K^-$ mesons with quark content $\bar{u}s$ off a target of protons with quark content $uud$ we have a system with quark content $uud\bar{u}s$. This 5-quark configuration has the same quantum numbers as $uds$, i.e. of a $\Lambda$. Indeed, the well known $\Lambda(1520)$ is generated in such a way through s-channel resonance formation. If we scatter $K^+$ mesons with quark content $\bar{s}u$ from neutrons with quark content $udd$, we have a $uudd\bar{s}$ system. Such a system does not have flavor quantum numbers that can be represented by a 3-quark strange baryon. It would be a baryon with minimally five quarks, and a "flavor exotic". Can such a pentaquark baryon exist? QCD does not seem to present any restriction that would preclude such an object to exist. After decades of inconclusive searches, the interest in pentaquark baryons was rekindled with a paper by D. Diakonov, V. Petrov, and M. Polyakov in 1997~\cite{dpp97}, who predicted a narrow state with exotic flavor at a mass of 1530 MeV.  This state is now called the $\Theta^+(1540)$. The  prediction was made in the chiral soliton model ($\chi$SM). The $\Theta^+$ would be an isosinglet with strangeness = +1, in an anti-decuplet of 5-quark states with $J^P = \frac{1}{2}^+$, as shown in Fig.~\ref{fig:antidecuplet}. Only three of these states have exotic quantum numbers, the $\Theta^+$, and the $\Xi^{--}$ with quark content $ddss\bar{u}$, and the $\Xi^+$ with quark content $uuss\bar{d}$. The other members of the anti-decuplet have quantum numbers that are also observed for 3-quark states. The predicted low mass and the narrow width made the $\Theta^+$ an interesting object for experimentalists to search for. 
Evidence for and against the existence of pentaquarks are discussed in the talk by T. Nakano\cite{nakano04}.
I will restrict myself to aspects of the pentaquark program with the CLAS detector\cite{clas_d,clas_p,g10,g11,g12,eg3}. 

The CLAS results on deuterium\cite{clas_d} are well known and the experiment has been repeated with higher luminosity 
and is currently being analyzed.  The results on the proton have only been published earlier this year, however, and are not as widely known as some of the other results. They are interesting as they may present hints at the $\Theta^+$ production mechanism at higher energies\cite{karliner04}. 
\begin{figure}[tbhp]
\vspace{5.5cm}
\centering{\includegraphics{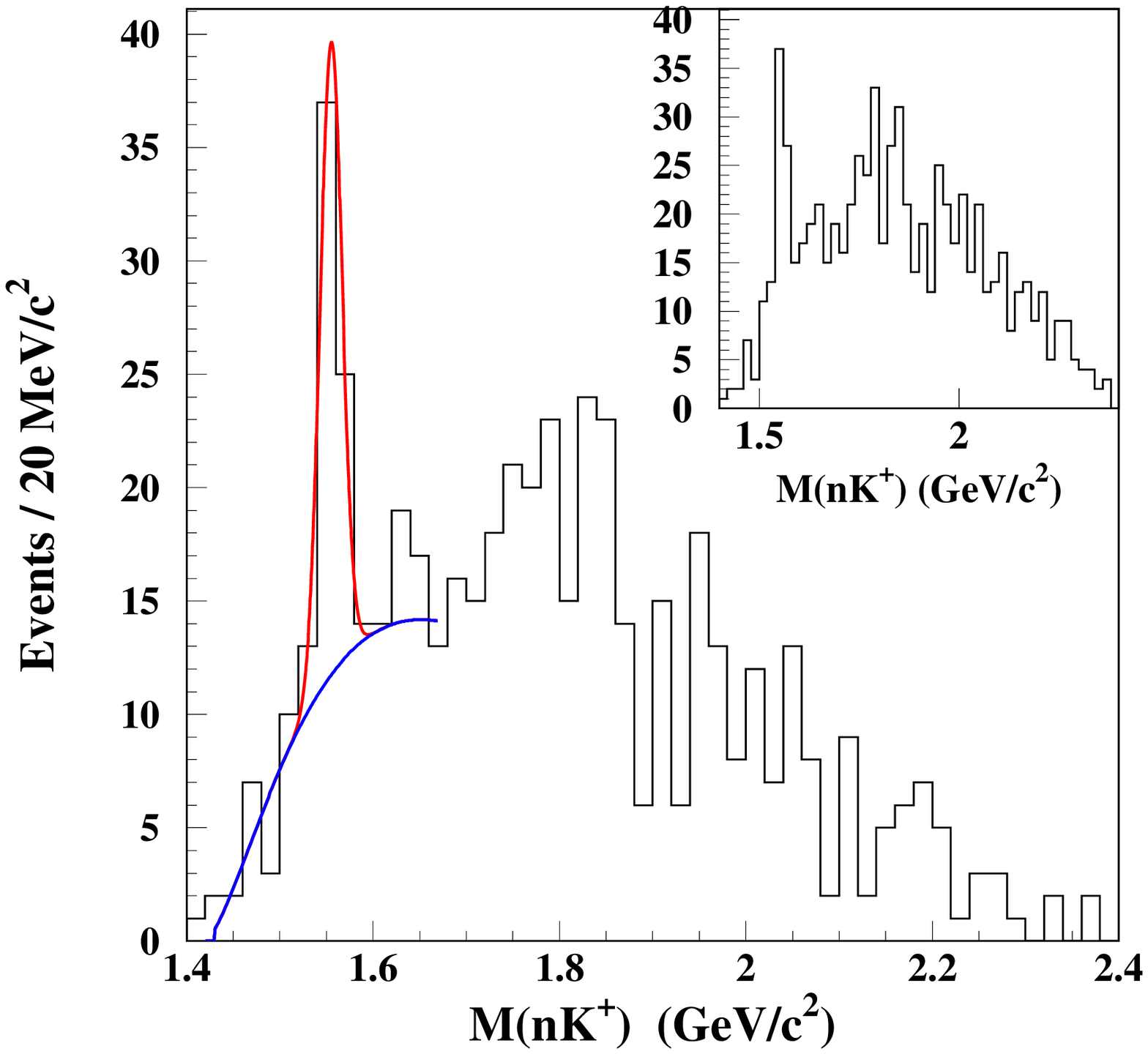}}
\centering{\includegraphics{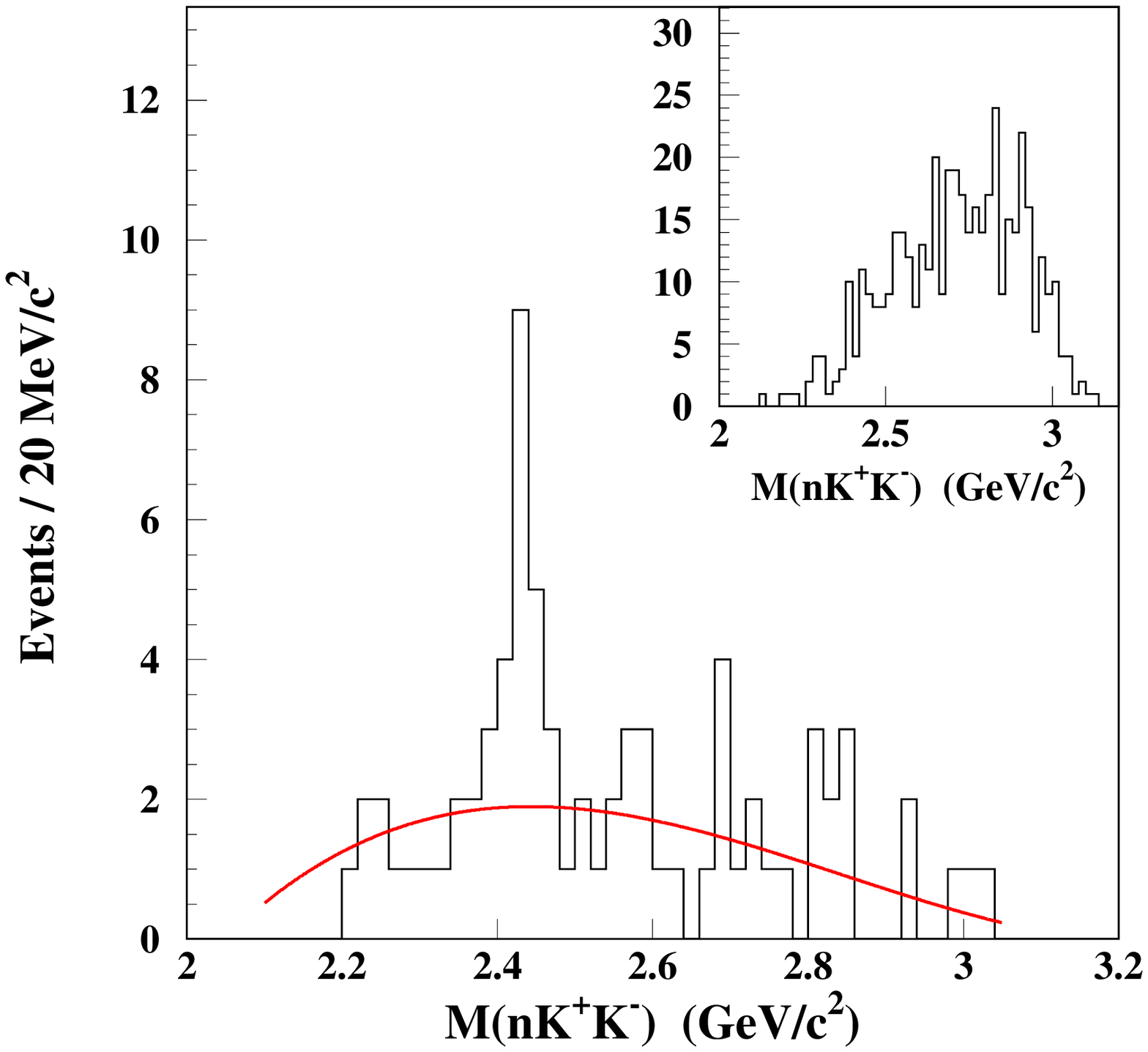}}
\caption{Left: Invariant mass distribution of $M(nK^+)$ after all cuts. The inset shows the $nK^+$ 
mass distribution with only the $\cos\theta^*_{\pi^+} > 0.8$ cut applied. Right: Mass distribution $M(K^-nK^+)$ for events selected in the peak region of the graph on the left. The inset shows the distribution for events outside of the $\Theta^+$ region.} 
\label{fig:clas_p}
\label{fig:clasp_N*}
\end{figure}
In this experiment, the process $\gamma p \rightarrow \pi^+K^+K^-n$ was measured at photon energies from 3 to 5.4 GeV. 
Events are selected with the $\pi^+$ at forward angles, and with the $K^+$ at large angles. The rationale for this selection is that the $\Theta^+$ may be produced through the decay of an intermediate $N^*$ resonance that is excited through a $\pi^-$ exchange with the proton, i.e. $\pi^- p \to N^* \to \Theta^+ K^-$.  The event selection will reduce t-channel contributions to $K^+$ production at forward angles. The mass spectrum is shown in Fig.~\ref{fig:clas_p}. A significant peak is seen at a mass of 1555 $\pm$ 10 MeV which has been associated with the $\Theta^+$.  In this analysis, background processes were subjected to a partial wave analysis allowing for a precise determination of the background shape under the $\Theta^+$ peak. The assumption of intermediate $N^*$ excitation can be tested by selecting the events in the region of the $\Theta^+$ and plotting the $nK^+K^-$ mass distribution, which is shown in Fig.~\ref{fig:clasp_N*}.  While the data hint at a narrow structure near a mass of 2.4~GeV, the statistics are too poor to allow more definite conclusions. Further studies with higher statistics are clearly needed to get a better handle on possible production mechanisms of the $\Theta^+$ at different kinematics.

\subsection{Search for the $\Theta^+$ and excited states at Jefferson Lab.}

The existing data show the capabilities of CLAS to select
exclusive final states with high multiplicity. The reaction channels are 
cleanly identified with small background due to misidentified particles. 
Concurrent reactions decaying to the same final states were seen and rejected from the 
final event sample. However the number of events in the $\Theta^+$ peak is rather small
with sizeable background, and does not allow to perform detailed checks of systematic dependencies.
\begin{figure}
\vspace{6.cm}
\includegraphics{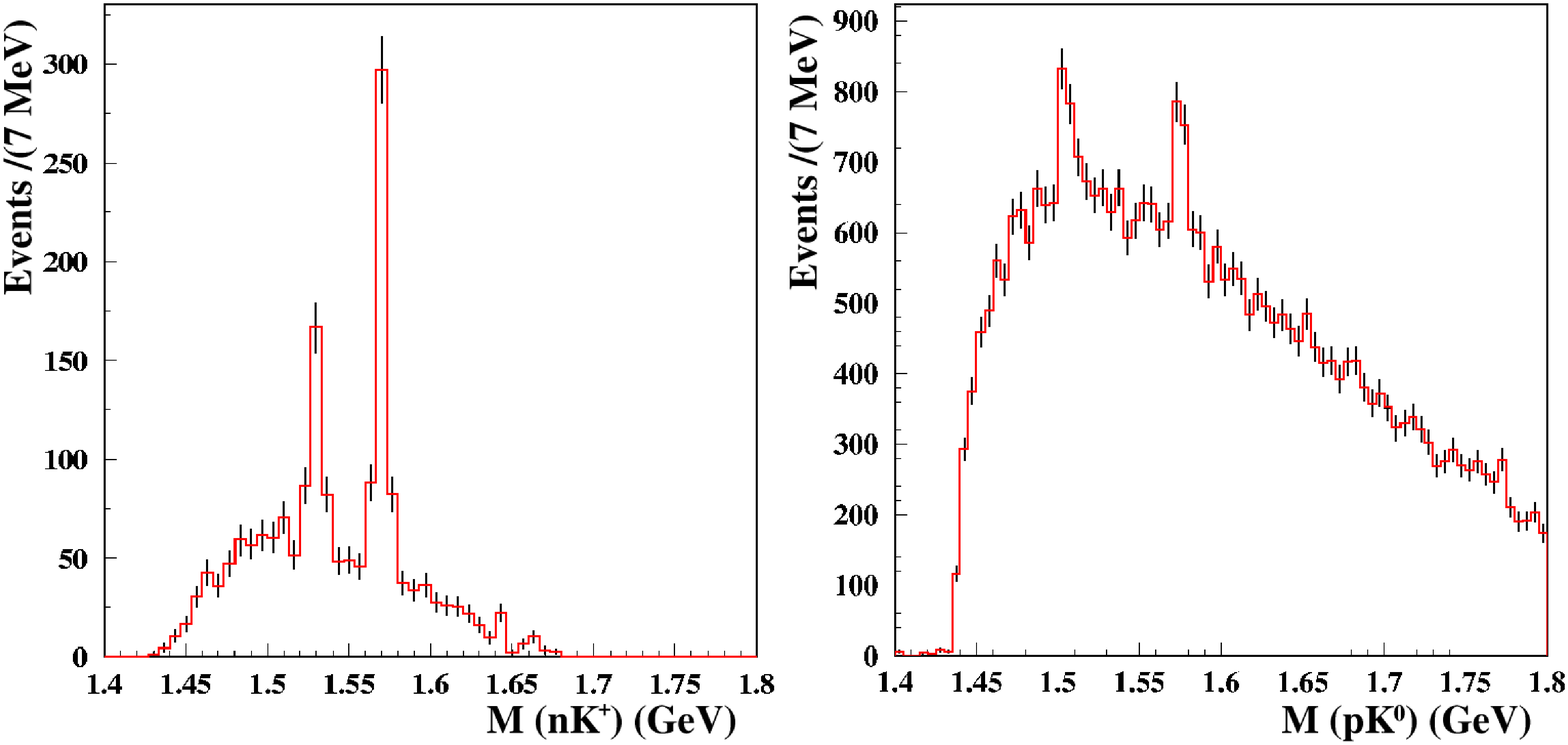}
\caption[]{Expected statistical accuracy of the mass spectra for the reactions 
$\gamma p \to \Theta^+ (\Theta^{+*}) \bar K^0$, with $\Theta^+ (\Theta^{+*})$ 
decaying into $K^+n$ (left) and $pK^0$ (right). A production cross section of 10nb is assumed.}
\label{fig:exp_masses}
\end{figure}
To obtain a more definitive result for or against the existence of pentaquark states, four dedicated 
experiments are underway using CLAS at Jefferson Lab. 
The goals and experimental conditions of these experiments are summarized in Table \ref{table:newruns}.
\begin{table}
\tbl{New experiments proposed in Hall B for the search of pentaquark states .}
{\begin{tabular}{|c|c|c|c|l|l|}
\hline
{Run} & {Beam} & {Energy} & {Target} & {Reaction} & {Status}\\ 
\hline\hline
{g10} & {$\gamma$} & {3.8 GeV} & {LD$_2$} & {$\gamma d \to \Theta^+ K^- p$} & {Completed}\\ 
{} & {} & {} & {} & {$\gamma d \to \Theta^+ \Lambda^0$} & {}\\ 
\hline
{g11} & {$\gamma$} & {4.0 GeV} & {LH$_2$} & {$\gamma p \to \Theta^+ \bar K^0$} & {Completed}\\ 
{} & {} & {} & {} & {$\gamma p \to \Theta^+ K^- \pi^+$} & {}\\ 
\hline
{eg3} & {$\gamma$} & {5.75 GeV} & {LH$_2$} & {$\gamma p \to \Xi^{--} X$} & {In progress}\\ 
{} & {} & {} & {} & {$\gamma p \to \Xi^{+} X$} & {}\\ 
\hline
{g12} & {$\gamma$} & {6 GeV} & {LH$_2$} & {$\gamma p \to \Theta^+ K^- \pi^+$} & { To be scheduled}\\ 
{} & {} & {} & {} & {$\gamma p \to \Theta^+ \bar K^0$} & {}\\
{} & {} & {} & {} & {$\gamma p \to K^+K^-\Xi^{-}$} & {}\\ 
\hline
\end{tabular}
\label{table:newruns}}
\end{table}
The g10 experiment\cite{g10}, which has taken data during the spring of 2004, aims at studying production channels $\gamma d \to pK^-\Theta^+$, $\gamma d \to p K^0 X$, and $\gamma d \to \Lambda \Theta^+$ with an order of magnitude improved statistics over the previous g2a run. The g11 experiment studies $\gamma p \to \Theta^+ \bar K^0$ and $\gamma p \to \Theta^+ K^- \pi^+$, and two decay modes, $\Theta^+ \to nK^+$ and $\Theta^+ \to p K^0$, increasing by an order of magnitude the statistics of the previous data. Both experiments have similar experimental setups and beam conditions as used in previous runs. 

If the existence of the $\Theta^+$ can be established with certainty, the new data will allow us to make progress on establishing the phenomenology of the $\Theta^+$ spectrum, e.g. determining in what production channels the $\Theta^+$ is seen and what higher mass states are excited. The expected statistical accuracy is shown in Fig.~\ref{fig:exp_masses}, where the background was estimated based on the existing data and the signal was simulated assuming a production cross section of $\sim 10$ nb. If the existence of the $\Theta^+$ is confirmed or new states are seen, these data will provide accurate measurements of the mass. In addition,
the large acceptance of CLAS will allow us to measure both the production and decay angular distributions, providing information on the production mechanism and spin.

\subsection{Search for $\Xi^{--}$ and $\Xi^-$ baryons.} 

The anti-decuplet of 5-quark states also contains $\Xi$ baryons, 
two of them of exotic nature, the $\Xi^{--}$ and the $\Xi^{+}$. Evidence for 
such states has so far been seen in only one experiment\cite{na49}. Two new experiments with 
CLAS are in preparation to search for $\Xi$ baryons in the anti-decuplet.  
The eg3 experiment\cite{eg3} uses an untagged photon beam of 5.75 GeV impinging 
on a liquid-deuterium target. The process $\gamma n \rightarrow \Xi^{--} X$ will be searched for by 
measuring the decay chain  $\Xi^{--} \rightarrow \pi^- \Xi^- \rightarrow  \pi^- \Lambda \rightarrow \pi^- p$. 
 One proton and three $\pi^-$ emerging from three different vertices  have to be reconstructed. 
The experiment is taking data in the winter 2004/2005. 
The second experiment\cite{g12} is part of g12. It uses the missing mass method 
to search for the  $\Xi^-$ in the exclusive reaction $\gamma p \rightarrow  K^+K^+X$.
If the NA49 results are confirmed, the $\Xi^-$ should be seen in the missing mass 
spectrum as a peak at 1862 MeV. The excellent missing mass resolution that is available with 
CLAS is required for such measurements\cite{price04}.

\end{document}